\documentclass[12pt]{article}
\usepackage{cite,epsfig,amssymb,amsmath,graphicx,color}
\newcommand{\be}{\begin{equation}}
\newcommand{\ee}{\end{equation}}
\newcommand{\bear}{\begin{eqnarray}}
\newcommand{\eear}{\end{eqnarray}}
\newcommand{\ba}{\begin{array}}
\newcommand{\ea}{\end{array}}
\newcommand{\nn}{\nonumber}

\usepackage[normalem]{ulem}
\usepackage{varioref}

\begin{document}

\begin{center}
{{{\Large \bf  Large Scale Suppression of Scalar Power
\\ on a Spatial Condensation}}
\\ [17mm]
Seyen Kouwn$^{1,2}$,~~O-Kab Kwon$^{2,3}$, and ~~Phillial Oh$^{4}$\\[3mm]
{\it$^{1}$Korea Astronomy and Space Science Institute,
Daejeon 305-348, Republic of Korea\\
$^{2}$Institute for the Early Universe and Department of Physics, \\ Ewha Womans University,
Seoul 120-750, South Korea \\
$^{3}$Department of Physics, Kyungpook National University, Taegu 702-701, Korea
$^{4}$Department of Physics,~BK21 Physics Research Division,
~Institute of Basic Science,\\
Sungkyunkwan University, Suwon 440-746, Korea}\\[2mm]
{\tt seyenkouwn@kasi.re.kr,~okabkwon@ewha.ac.kr,~ploh@skku.edu} }
\end{center}

\vspace{10mm}

\begin{abstract}

We consider a deformed single field inflation
model in terms of three SO(3) symmetric moduli fields.
We  find that spatially linear solutions for the moduli
fields induce a phase transition during the early stage of the inflation
and the suppression of scalar power spectrum at large scales. 
This suppression can be an origin of 
anomalies for large scale perturbation modes in the cosmological observation.   
\end{abstract}
\newpage

\tableofcontents

\section{Introduction}

Recent measurements of the cosmic microwave background 
by the WMAP and Planck collaborations~\cite{Ade:2013zuv,WMAP9} support 
the inflationary scenario. Most of inflationary models predict 
a nearly Gaussian and scale invariant power spectrum 
of adiabatic perturbation modes, which can be realized by the single field slow-roll
inflationary model. 
However, the most recent data released by Planck collaboration~\cite{Ade:2013zuv} 
reported a statistically significant anomalies at low multipoles, 
which corresponds a power deficit 5 - 10$\%$ at multipole range $l\lesssim 40$ 
with $2.5$ - $3\sigma$ level. 
Therefore, the usual single field inflation model needs some modification 
to express the large scale scalar power suppression.  
This suppression of the scalar power spectrum was then used to 
explain the observed low quadrupole in the CMB anisotropy.

There are several interesting approaches which address the scalar power suppression and are relevant 
to the suppression method pursued in this paper. 
The first one is the mechanism studied by Hazra et al~\cite{Hazra:2014jka}, 
where the authors introduced   
a steep potential during the first few e-foldings of inflation.  
Then there appears a fast roll phase during the large scale modes cross the horizon and 
the resulting scalar power spectrum is suppressed since it is inversely proportional 
to the inflaton velocity. See also \cite{Joy:2007na,Hazra:2014goa}.
The second one is related to a nonzero spatial curvature 
in a single field inflation model. In this case one can also induce the suppression of 
the scalar power spectrum on large scales~\cite{Lyth:1990dh,Ratra:1994vw}.
In the same line of thought,  recently White et al revisited  the open inflation 
model~\cite{White:2014aua}, which gives rise to a suppression of the scalar power 
on large scales. Here the main source of the suppression is also the steepening of the potential 
due to the barrier that separates the true and false vacua. 
 
In this paper we consider a modification of the canonical single field inflation model,
which induces large negative running of $n_s$ and results in a suppression
of scalar power spectrum on large 
scales.\footnote{The large negative running of $n_s$ can be introduced 
in the context of reconciling the results of Planck 
and BICEP2 collaborations~\cite{Ade:2014xna}, though 
it has been pointed out that uncertainty from the foreground effect 
can dominate the excess~~\cite{Mortonson:2014bja, Adam:2014bub, Cheng:2014pxa} observed by the BICEP2 collaborations. See also for the suppression of the scalar power spectra on large scales~\cite{Contaldi:2014zua,Hazra:2014aea,Hazra:2014jka,Bousso:2014jca,White:2014aua} after the BICEP2 observation.}
As a specific model, we consider a deformation of a single field inflation model
by adding kinetic terms for a number of scalar moduli fields,
\begin{align}\label{modfld}
-\frac12 \int d^4 x \sqrt{-g}\, \sum_{m=1}^{\bar N} \partial_\mu \sigma^m \partial^\mu \sigma^m.
\end{align}
We also consider a background solution with spatially linear configurations,
$\sigma^a\sim x^a$, $(a=1,2,3)$, and $\sigma^i=0$, ($i=4,\cdots, {\bar N}$).
Then the usual cosmological evolution for the single field under the FRW metric with the background
solution for $\sigma^a$ guarantees the homogeneity and isotropy of the cosmological
principle~\cite{ArmendarizPicon:2007nr,Lee:2009zv,Endlich:2012pz,Koh:2013msa}  .
In the perturbation level, fluctuations for $\sigma^i$, ($i=4,\cdots, {\bar N}$), are decoupled and have
no influence to cosmological observables~\cite{Koh:2013msa}.
For this reason, we consider the ${\bar N}=3$ case for simplicity.
This model corresponds to the case with $f(\varphi)=1$ in the work~\cite{Koh:2013msa}.
On the other hand, without the usual single inflaton field contribution, inflation is also possible when one uses higher order combination of  $ X=\partial_\mu \sigma^a \partial^\mu \sigma^a$, ($a=1,2,3$),  with spatially linear configuration of $\sigma^a$.
This inflation model is known as the solid inflation~\cite{Endlich:2012pz}.
See also \cite{Bartolo:2013msa}.

In our model the background evolution  is the same with that of the single field
model with the curvature term of the open universe.
That is, the solution $\sigma^a\sim x^a$ induces the curvature term of the open universe in the Friedmann equation,
though we start from the flat FRW metric.\footnote{We call the remnant of the  solution
$\sigma^a\sim x^a$ in the background evolution as {\it spatial condensation}.}
The curvature term is proportional to inverse square of the scale factor,
and so the effect of {\it the  spatial condensation}
appears during the very early stage of the inflation
and disappears quickly as the scale factor grows up. Since we start from the phase
where the curvature term is much more dominant than the potential term of the single field,
there appears a phase transition  from the curvature term dominant phase to the potential term dominant phase.
Due to the phase transition in the early stage of the background evolution, there appears the suppression of the scalar power spectrum.  This situation has some resemblance
to that of inflation models referred as `whipped inflation'~\cite{Hazra:2014aea,Hazra:2014jka}
and `open inflation'~\cite{Linde:1998iw,Bousso:2014jca,White:2014aua},
in which there exist phase transitions from the fast-roll phase to the slow-roll phase
of  the single scalar field model.   These phase transitions during the early stage of inflation induces
the suppression of the scalar power spectrum on large scales, though the detailed suppression mechanisms are different from that in our model.

The organisation of this paper is as follows. In the next section, we explain
the properties of the  background evolution under  the spatial condensation.
In section 3, we investigate the effects of the spatial condensation
in linear perturbation level.
We find the suppression of the scalar power spectrum and large value of the running of the scalar
spectral index on large scales. We conclude in section 4.

\section{Background Evolution on a Spatial Condensation}
We start from the action for the single field inflation model with an additional
triad of moduli scalar fields,
\begin{align}\label{act1}
S = \int d^4x \sqrt{-g}\Biggr[ {M_{{\rm P}}^2\over 2}R -{1\over2}\,
\partial_\mu \varphi \partial^\mu \varphi -V(\varphi)
-{1\over2}  \partial_\mu \sigma^a \partial^\mu \sigma^a  \Biggr],
\end{align}
where $a=1,2,3$ and  $M_{{\rm P}}$ denotes the Planck mass,
$M_{{\rm P}} \equiv (8\pi G)^{-1/2}$.
The SO(3)-symmetric fields $\sigma^a$ have no potential.
Then equations of motion of the scalar fields $\sigma^a$ are read as
\begin{align}\label{sigmaEOM}
{1\over\sqrt{-g}} \partial_\mu \Big( \sqrt{-g} g^{\mu\nu}\partial_\nu \sigma^a \Big)=0.
\end{align}
Under the background FRW metric,
$ds^2 = -dt^2 + a(t)^2(dx^2 + dy^2 + dz^2)$
with the scale factor $a(t)$,  the spatially linear configuration
\begin{align}\label{LinSig}
\sigma^a = M_{\rm P}^2\alpha\, x^a
\end{align}
satisfies the equations \eqref{sigmaEOM}.
Here the constant gradient $\alpha$ is an arbitrary dimensionless parameter.
As usual inflation models which are not compatible with the cosmological
principle of homogeneity and isotropy in the background evolution,
we assume that the field $\varphi$ depends on time only.
Then the remaining equations of motion of $g_{\mu\nu}$ and $\varphi$  in \eqref{act1}
are given by
\begin{align}
&H^2 = \frac{1}{3M_{\rm P}^2}\left( {1\over 2}\dot{\varphi}^2 + { 3 M_{\rm p}^4
\alpha^2 \over 2a^2 } + V  \right),
\nn \\
&\dot{H} = -{1\over 2M_{\rm P}^2}\left(\dot{\varphi}^2 +  {M_{\rm p}^4
\alpha^2 \over  a^2 }   \right),
\label{backEOM}  \\
&\ddot{\varphi} +3 H\dot{\varphi}
+ V_\varphi= 0 \,, \nn
\end{align}
where  $H\equiv \dot{a}/a$ and $V_\varphi\equiv dV/d\varphi$.
As was discussed in \cite{Koh:2013msa},  the $\alpha$-terms
in \eqref{backEOM} correspond to the curvature terms by identifying the curvature
constant $K$ as $K = - M_{\rm P}^2\alpha^2/2$. Since the curvature constant is negative
in this case, the equations representing the background evolution in \eqref{backEOM}
are the same with those of the open universe in the single field inflation model.
Though the single field model in the open universe is the same with our model in
the background level, they are different in the perturbation level due to the contribution of fluctuation modes of $\sigma^a$.
In our case three degrees of freedom (one scalar mode and two vector modes)
originated from the triad of scalar fields
appear in the perturbation level, while there is no additional perturbation degree of freedom in the usual single field inflation model with negative curvature constant.

Now we investigate some characteristic properties of the background evolution
of our model.
The effect of $\alpha$-terms
in \eqref{backEOM} is decreasing rapidly
during the inflation and has some influence on the early time of the inflation period.
Especially as we see in the first line of \eqref{backEOM}
the Hubble horizon $r_H\equiv 1/H$ starts from a small value when
we introduce a  large value of the $\alpha$-dependent term at the initial state,
increases during the early
stage of the inflation, and approaches the value of the single field inflation model at late time.

In our model,  the suppression of scalar power spectrum on large scales can be  achieved by introducing a large value of $\alpha$-term
at the early time of inflation. For that purpose,
we consider the case that
the $\alpha$-term in the first equation of \eqref{backEOM} is much larger than
the potential term of the inflaton field, i.e.,
\begin{align}\label{inival}
{ 3 M_{\rm p}^4 \alpha^2 \over 2a^2 } \gg V (\varphi).
\end{align}
Obtaining analytic solution for the equations in \eqref{backEOM} is a formidable task
for the potentials of the large field inflation models.
So we rely on a semi-analytic way to figure out the behaviour of the background evolution governed by the equations  in \eqref{backEOM},
based on numerical method.
By employing the simplest scalar potential $V(\varphi) = \frac12 m^2\varphi^2$, for
concreteness, we find that the scalar field $\varphi$ remains almost constant until
the e-folding number $N = 10\sim 20$. See Fig.1.
Then there appears a stage that
the value of $\alpha$-term is comparable to that of the potential, i.e.,
\begin{align}
{ 3 M_{\rm p}^4 \alpha^2 \over 2a^2 } \simeq V (\varphi).
\end{align}
After the universe passes through this stage,
the scalar field starts to decrease and follows the behaviour of the canonical
slow-roll inflation.
The behaviours of the background scalar field and the Hubble horizon with respect to
the e-folding number $N$ are plotted in Fig.1.
\begin{figure*}[ht]\label{nsolbk}
\begin{center}
\scalebox{0.57}[0.57]{\includegraphics{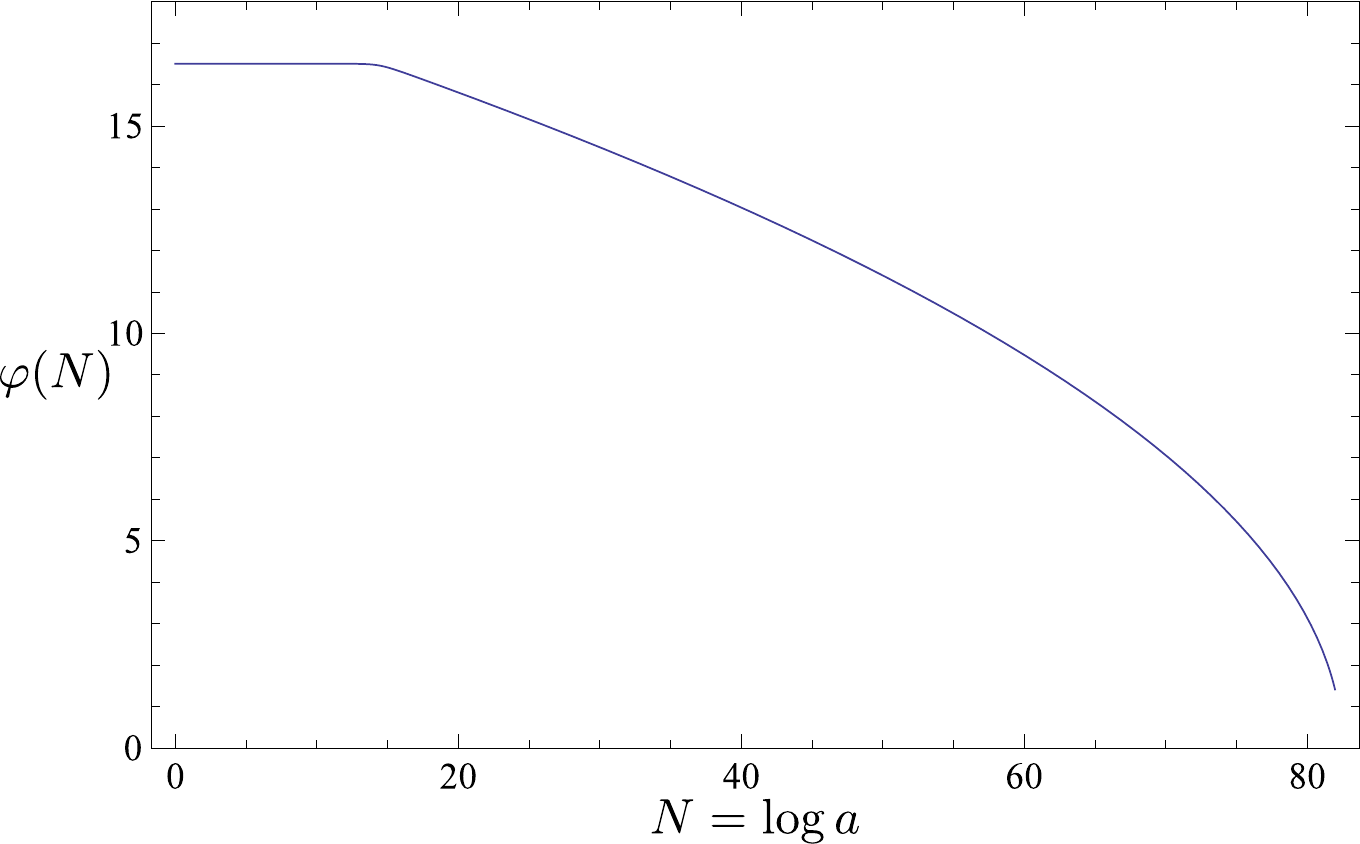}~~~~~~~\includegraphics{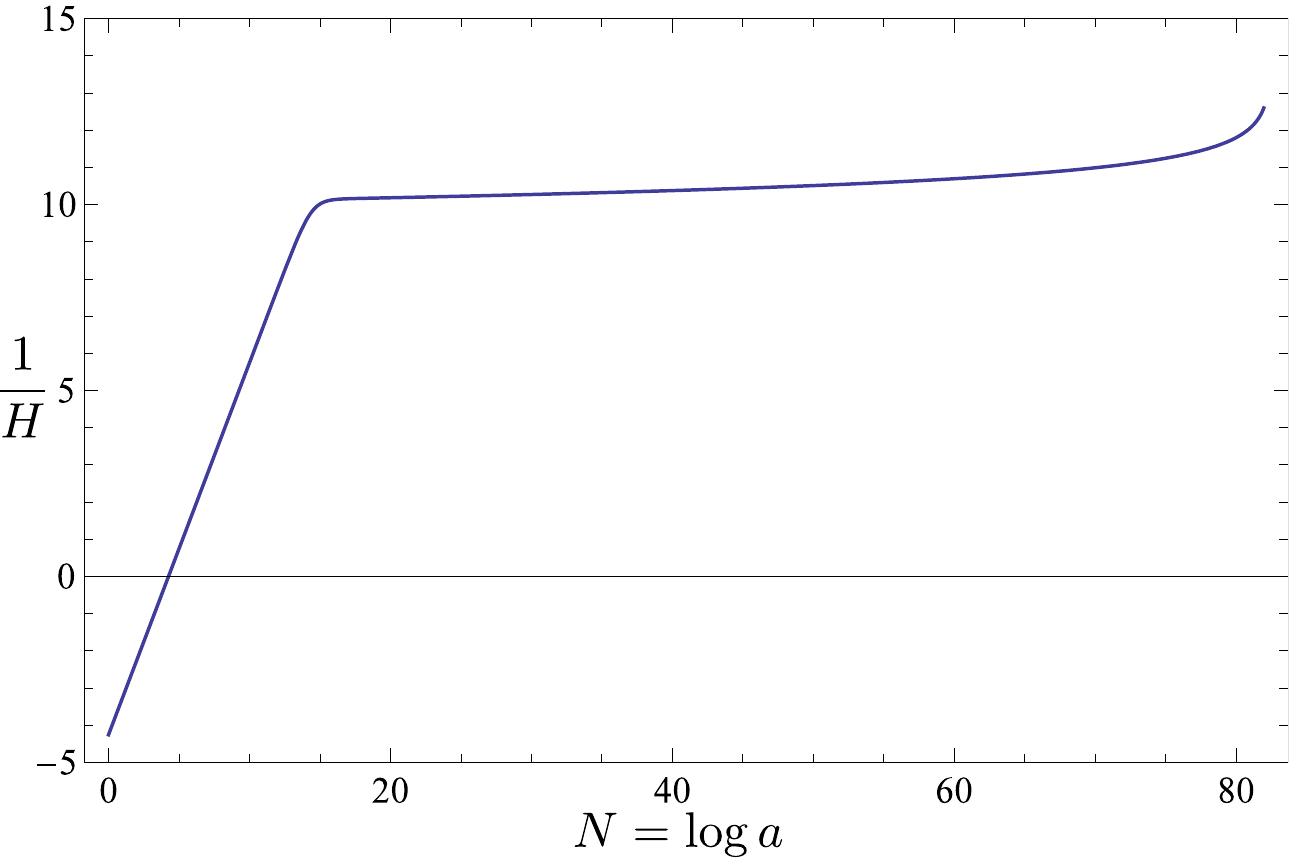}}\label{Fig1}
\end{center}
\caption{\small
The graphs of  inflaton field $\varphi(N)$ (left) and Hubble radius $1/H(N)$~(right).
We set the pivot scale $ k_0 = 0.05 {\rm Mpc}^{-1}$ and
$N_* = 60$.
We choose the initial condition as $a_i = 1$ , $\varphi_i = 16.5 M_{{\rm P}}$,
$\dot{\varphi_i} = 0$, $m = 5.85 \times 10^{-6} M_{{\rm P}}$,  and $\alpha = 10^2$.
}
\end{figure*}

As we see in Fig.1, there is a sharp transition point near
\begin{align}\label{Neq}
N_{{\rm eq}} \simeq \ln \sqrt{\frac{3  \alpha^2M_{\rm P}^4}{2 V\big(\varphi (N_{{\rm eq}})\big)}},
\end{align}
where $N_{{\rm eq}}\equiv \log a_{{\rm eq}}$ represents the e-folding number
when the $\alpha$-term is the same with the potential of the scalar field.
After the transition point the background evolution rapidly follows the behaviour
of the usual slow-roll inflation by rolling down the potential slop.
For this reason, we can approximate the background equations in \eqref{backEOM}
under the assumption of the slow-rolling of the scalar field as follows:
\begin{align}
{\rm early~time} ~:~ &3H^2 \simeq  {3\alpha^2 M_{{\rm P}}^2 \over 2 a^2}
+ \frac{\Lambda}{M_{{\rm P}}^2} \,,
\label{bkseq1} \\
{\rm late~time} ~:~ &3H^2 \simeq {3\alpha^2 M_{{\rm P}}^2\over 2 a^2}
+\frac{ V(\varphi)}{M_{{\rm P}}^2} \,, \label{bkseq2}
\end{align}
where $\Lambda \equiv V(\varphi_i)$ with an initial value of the scalar field
$\varphi_i$. From the relation \eqref{Neq} we also have the relation
\begin{align}\label{Neq2}
N_{{\rm eq}}  \simeq \ln \sqrt{\frac{3  \alpha^2M_{\rm P}^4}{2 \Lambda}}.
\end{align}

In the early time, the background equation \eqref{bkseq1}  has a solution~\cite{Masso:2006gv,Yamamoto:1995sw}.
\begin{align}
a(t) \simeq a_{eq}\, \sinh \left(\sqrt{\frac{\Lambda}{3M_{{\rm P}}^2}}(t+ t_0)\right),
\end{align}
where $t_0$ is the initial time with the scale factor $a_0$.
 On the other hand, in the late time for a given value of $\alpha$,
 the scale factor `$a$' is already very large,
 and then the $\alpha$-term in \eqref{bkseq2} becomes much smaller
 than the potential term.
 Based on the numerical result during the late time in Fig.1, we see that
 the scalar field starts to roll down
 the potential slope matching the behaviour of the usual slow-roll approximation.
 Since the behaviour of the background evolution for the case
 $\frac{3\alpha^2 M_{{\rm P}}^2}{2 a^2} \ll \frac{V}{M_{{\rm P}}^2}$
 was already investigated in \cite{Koh:2013msa}, we omit the detailed
 background behaviours in this paper.

\section{Suppression of Large Scale Scalar Power Spectrum}

\subsection{Generality }

We consider the linear scalar perturbation of the FRW metric,
\begin{align}\label{met_pert}
ds^2 = -\left(1+2A\right)dt^2 +2a\, \partial_iB\,dt dx^i
+a^2\Big[\left(1-2\psi\right)\delta_{ij}+2\partial_i\partial_jE \Big]dx^i dx^j,
\end{align}
where $A$, $B$, $\psi$, and $E$ are four scalar modes.
In the linear perturbation, there is also a contribution from fluctuations of scalar fields,
\begin{align}
\varphi(t,x) &= \varphi(t) + \delta\varphi(t,x) \,,
\nn \\
\sigma^a(t,x) &= \sigma^a(x) + \delta\sigma^a(t,x) \,.
\label{phisig}
\end{align}
The three perturbation modes $\delta \sigma^a$ are decomposed into
one scalar and two vector modes,
\begin{align}
\delta\sigma^a = \delta\sigma^a_{\parallel} + \delta\sigma^a_\perp.
\end{align}
In this work, we focus on  the scalar mode $\delta\sigma^a_{\parallel}$
in perturbation level.\footnote{In the linear perturbation level, the two vector modes $\delta^a_{\perp}$ have no contribution to the scalar mode~\cite{Koh:2013msa}. 
To see this fact explicitly, 
one can read the $(0,i)$-component of the perturbed Einstein 
equation $\delta G^0_{~i} = M_{\rm P}^{-2} \,\delta T^0_{~i}$ as
\begin{align}\label{G0i}
-2\,\partial_i\left(HA+\dot{\psi}\right)
= M_{\rm P}^{-2}\delta T^0_{~i}
= M_{\rm P}^{-2}\left(-\partial_i(\dot\varphi\delta\varphi)
-\alpha  \delta\dot\sigma^i + \frac{\alpha^2}{a}\partial_i B\right).
\end{align}
Taking the curl of both sides of \eqref{G0i}, we obtain
$\epsilon_{ijk}\partial^j \delta\sigma^k=0$. That is, only the longitudinal scalar mode can satisfy this relation.  }
Since the kinetic term of $\sigma^a$ in \eqref{act1} is a well-defined quadratic form, the kinetic term of 
the three perturbation modes $\delta\sigma^a$'s is also well-defined and does not violate the energy condition. 
Here we express the longitudinal mode $\delta\sigma^a_{\parallel}$ in terms of a scalar mode $u$ with a
normalization~\cite{ArmendarizPicon:2007nr,Koh:2013msa},
\begin{align}
\delta\sigma^a_{\parallel}=\frac1{k}\partial_a u,
\end{align}
where $k$ is the comoving wave number. 
We introduce the $1/k$ factor to define the canonical kinetic term for the scalar mode $u$ in the form 
of the perturbed Lagrangian.

Employing the spatially flat gauge
$(\psi=0\,\&\, E=0)$, the perturbed scalar equations are
reduced to
\begin{align}\label{sflat}
&\ddot{Q}_\varphi+3H\dot{Q}_\varphi
+\Biggr( {k^2\over a^2}+{\dot{\varphi}V_\varphi\over M_{\rm P}^2 H}+V_{\varphi\varphi}
 \Biggr)Q_{\varphi}
+2\Biggr({\dot{H}\dot{\varphi}\over H}-\ddot{\varphi}\Biggr)A=0,
\nn \\
&\ddot{Q}_{u}+3H\dot{Q}_u
+\Biggr({k^2\over a^2}
+ {2\alpha^2 M_{\rm p}^2 \over a^2}\Biggr)Q_u=0,
\end{align}
where  $Q_\varphi \equiv \delta\varphi -\frac{\dot\varphi}{H}\psi$ and $Q_u = u
-\alpha k M_{{\rm P}}^2 E$ are gauge invariant quantities~\cite{Koh:2013msa}.
Scalar modes $A$ and $B$ satisfy the constraints,
\begin{align}\label{cAB}
&3AH^2-{k^2BH\over a}=\frac1{2M_{\rm p}^2}\Big(A\dot{\varphi}^2-\dot{\varphi}
\dot{Q}_\varphi - V_\varphi Q_\varphi\Big) + \frac{\alpha k}{2 a^2}\, Q_u,
\nn \\
&2AH={\dot{\varphi} Q_\varphi\over M_{\rm p}^2}
+\frac{\alpha}{k}Q_u-{\alpha^2 M_{\rm p}^2 B\over a }.
\end{align}
Using these constraints one can express the modes $A$ and $B$ in terms of $Q_\varphi$ and $Q_u$. In this multifield perturbation system,
the comoving curvature perturbation is written as~\cite{Koh:2013msa}
\begin{align}
{\cal R}=H\Biggr[{
\dot{\varphi}Q_\varphi
-\alpha M_{\rm P}^2
\big({\alpha M_{\rm P}^2 B\over a}-{{\dot  Q_u }\over k}\big)
\over {\dot\varphi}^2
+{\alpha^2M_{\rm P}^4\over a^2}} \Biggr].
\label{ICP}
\end{align}
Differently from the single field inflation model, there is also non-vanishing
isocurvature perturbation~\cite{Koh:2013msa}. However, here we only concentrate on the adiabatic curvature perturbation.

\subsection{Suppression of the scalar power spectrum}

As we discussed in section 2,
there are two inflation phases in the background evolution,
the $\alpha$-term dominant phase
and the scalar potential dominant phase.
Due to the phase transition during the inflation, the computation of the power spectrum is different from
that of the usual single scalar field model.
In order to calculate  power spectrum and related observational quantities, such as
the scalar spectral index $n_s$ and the running of the spectral index $\alpha_s$,
we use the method developed in \cite{Joy:2007na}, where the authors calculated
the power spectrum of the single scalar field model with the potential having a step transition.
Due to the shape of the scalar potential, there are two inflationary phases, fast-roll phase and slow-roll phase.
The origin of the phase transition in \cite{Joy:2007na} is different from ours,
but there is a robust similarity between
these two cases in the sense that there is a transition during the inflation and the background
evolution approaches the usual slow-roll inflation phase of the single field model at late time.
For this reason, we follow the method developed in \cite{Joy:2007na} to compute the power spectrum and related
perturbation quantities.
Due to the phase transition in the background level, there is also a phase transition
to the perturbed equations in \eqref{sflat}.
We try to solve the perturbed equations for the $\alpha$-term dominant phase  and the scalar potential dominant phase  separately
and apply the matching condition at the transition point.

\subsubsection{Early time}

In the early time having the limit ${ 3 M_{\rm p}^4 \alpha^2 \over 2a^2 } \gg V $,
we obtain $A$, $B$ from \eqref{cAB} in the leading order of the limit as
\begin{align}\label{earlyAB}
A \approx \frac{\frac{\alpha}{k}M_{{\rm P}} Q_u}{ 2 + 3
\frac{\alpha^2M_{{\rm P}}^2}{k^2}},
\qquad
B \approx - \frac{\sqrt{\frac{4 V_0}{3\alpha^2M_{{\rm P}}^4}}\,
Q_u}{ zM_{{\rm P}}^2\left(2 + 3 \frac{\alpha^2M_{{\rm P}}^2}{k^2}\right)}.
\end{align}
Using the relation \eqref{earlyAB} and the fact that
$\varphi$ is almost a constant during the $\alpha$-term dominant phase,
we conclude that the $A$-dependent term
in \eqref{sflat} is negligible.
One can also neglect $V_\varphi$ and $V_{\varphi\varphi}$-dependent terms in \eqref{sflat}
since $V(\varphi) $ is almost constant in the early time phase.
Introducing the Sasaki-Mukhanov variables,
\begin{align}
{\cal V}\equiv a Q_\varphi, \qquad {\cal U} \equiv a Q_u,
\end{align}
and the conformal time coordinate $\tau=\int dt/a$,
we obtain the decoupled differential equations for ${\cal V}$ and ${\cal U}$ as
\begin{align}\label{earlyVU}
&{\cal V}_e'' + \left(k_{e1}^2 - \alpha^2M_{{\rm P}}^2 \,{\rm csch}^2
\left( \frac{\alpha (-\tau)}{\sqrt{2}}\right)\right) {\cal V}_e =0,
\nn \\
&{\cal U}_e'' + \left(k_{e2}^2 - \alpha^2 M_{{\rm P}}^2\,{\rm csch}^2
\left( \frac{\alpha (-\tau)}{\sqrt{2}}\right)\right) {\cal U}_e =0,
\end{align}
where the prime represents the differentiation with respect to the conformal time, the subscript `$e$' denotes the early time phase,  and
\begin{align}
k_{e1}\equiv \sqrt{k^2 - \frac{\alpha^2M_{{\rm P}}^2}{2}}, \qquad
k_{e2} \equiv \sqrt{k^2 + \frac{3\alpha^2M_{{\rm P}}^2}{2}}\,.
\end{align}
Using general solutions for ${\cal V}_e$ and ${\cal U}_e$, we obtain
normalized solutions~\cite{Masso:2006gv},
\begin{align}\label{earlyVUsol2}
{\cal V}_e(\tau) &=  \frac{M_{{\rm P}}^{\frac32}}{\sqrt{2 k_{e1}}
\left(-\frac{\alpha M_{{\rm P}}}{\sqrt{2}} + i k_{e1}\right)}
\left(- \frac{\alpha M_{{\rm P}}}{\sqrt{2}}\, {\rm coth}
\left(\frac{\alpha(-\tau)}{\sqrt{2}}\right) + i k_{e1}\right) e^{-ik_{e1}\tau},
\nn \\
{\cal U}_e(\tau) &= \frac{M_{{\rm P}}^{\frac32}}{\sqrt{2 k_{e2}}
\left(-\frac{\alpha M_{{\rm P}}}{\sqrt{2}} + i k_{e2}\right)}
\left(- \frac{\alpha M_{{\rm P}}}{\sqrt{2}}\, {\rm coth}
\left(\frac{\alpha(-\tau)}{\sqrt{2}}\right) + i k_{e2}\right) e^{-ik_{e2}\tau}.
\end{align}
Here we adjusted the integration constants for the solutions ${\cal V}_e$ and
${\cal U}_e$ to get the  Bunch-Davis vacua in $\tau\to -\infty$ limit,
\begin{align}\label{earlyBDvac}
{\cal V}_e(\tau) = \frac{M_{{\rm P}}^{\frac32}}{\sqrt{2 k_{e1}}} \, e^{-i k_{e1}\tau},
\qquad
{\cal U}_e(\tau) = \frac{M_{{\rm P}}^{\frac32}}{\sqrt{2 k_{e2}}} \, e^{-i k_{e2}\tau}.
\end{align}

As we see in \eqref{earlyBDvac}, effective wave numbers for oscillation modes
${\cal V}_e$ and ${\cal U}_e$  at early time
are $k_{e1}$ and $k_{e2}$ which are deformed from the wave number $k$
due to the non-vanishing value of $\alpha$.
Then we find that there is minimum value of the comoving wave number $k$.
The modes below the minimum value always stay in super horizon scale and never
cross the horizon, so those modes do not contribute to current observable quantities.
Now we try to obtain the minimum value of the wave number.
As we will see later, the leading contribution to the power spectrum comes from
${\cal V}_e$ mode in the limit we are considering. So we focus
on the mode ${\cal V}_e$.
Horizon crossing condition for the mode ${\cal V}_e$ at  the early stage of inflation
is given by
\begin{align}
k_{e1} = a_* H_* \,,
\end{align}
and the corresponding conformal time $\tau_*$ is
\begin{align}
\tau_* = -{\sqrt{2} \over \alpha M_{{\rm P}}}\tanh^{-1}\left({\alpha M_{{\rm P}}
\over \sqrt{2} k_{e1}}\right) \,.
\end{align}
From this relation, we notice that at the early stage of inflation,
the Hubble crossing occurs only when perturbation modes satisfy the condition to give a real value of $\tau_*$,
\begin{align}
{\alpha M_{{\rm P}} \over \sqrt{2}\, k_{e1}} < 1 \,.
\end{align}
This condition determines the minimum value of
the comoving wave number,
\begin{align} \label{kminval}
k_{{\rm min}} = \alpha M_{{\rm P}} \,.
\end{align}
In the current observation for perturbation modes, the minimum comoving wave number is in the 
range $k_{{\rm min}}\lesssim 10^{-2} {\rm Mpc}^{-1}$. Due to this fact, one can roughly estimate 
the value of $\alpha$ as
\begin{align}
\alpha\sim \frac{{\rm Mpc}^{-1}}{M_{{\rm P}}} \ll 1. 
\end{align}

\subsubsection{Late time}

On the other hand, in the late time satisfying the condition
$\frac{3\alpha^2M_{{\rm P}}^4}{2 a^2}\, \ll\, V(\varphi) $,
one can express the scalar modes $A$ and $B$
in terms of the gauge invariant variables, $Q_\varphi$ and $Q_u$ from the constrain \eqref{cAB},
\begin{align}\label{ABapprox}
&A \simeq \frac{\dot\varphi}{2H M_{{\rm P}}^2}\, Q_\varphi + \frac{\alpha}{2kH}\, Q_u,
\nn \\
& B \simeq \frac{\alpha}{2k}\left(\frac{3 a}{k^2} - \frac1{aH}\right) Q_u + \frac{a\dot\varphi}{2Hk^2 M_{{\rm P}}^2}\, \dot Q_{\varphi}.
\end{align}
Using \eqref{ABapprox}, one can easily see that the coefficient of the $A$-dependent term
in the first line of \eqref{sflat}
is belonged to higher order  for slow-roll parameters.
For this reason, the differential equations for $Q_\varphi$ and $Q_u$ are decoupled
in the leading contribution of slow-roll parameters. Then we obtain differential equations
for ${\cal V}_l$ and ${\cal U}_l$ in the conformal time coordinate as
\begin{align}\label{VU}
&{\cal V}_{l}''+ \Biggr(\,k_{l1}^2 -{\mu_1^2
-\frac14\over \tau^2} \Biggr){\cal V}_{l} =0 ,\nn\\
&{\cal U}_{l}''+ \Biggr(\,k_{l2}^2 -{\mu_2^2
-\frac14 \over \tau^2} \Biggr){\cal U}_{l} =0,
\end{align}
where
the subscript `$l$' denotes the late time phase and
\begin{align}\label{wavevec}
k_{l1}^2 &\equiv k^2-{\alpha^2 M_{{\rm P}}^2\over6},
\quad
k_{l2}^2 \equiv k^2+{11\alpha^2 M_{{\rm P}}^2\over6},
\nn \\
\mu_1 &\simeq {3\over2}+3\epsilon-\eta,
\quad
\mu_2 \simeq {3\over2}+\epsilon.
\end{align}
Here the slow-roll parameters, $\epsilon$ and $\eta$, are defined as
\begin{align}
\epsilon ={\dot{\varphi}^2\over 2 M_{\rm P}^2 H^2}, \qquad
\eta ={V_{\varphi\varphi}\over 3H^2}.
\end{align}
General solutions of ${\cal V}_l$ and ${\cal U}_l$ modes for differential equations in \eqref{VU} are given by
 \begin{align}\label{laterVUsol}
 {\cal V}_{l} (\tau)&= M_{{\rm P}}^{\frac32} \sqrt{-\tau} \left(C_1 H^{(1)}_{\mu_1}
 (-k_{l1}\tau)
 + C_2 H^{(2)}_{\mu_1} (-k_{l1}\tau) \right),
 \nn  \\
 {\cal U}_{l}(\tau) &= M_{{\rm P}}^{\frac32} \sqrt{-\tau} \left( D_1 H^{(1)}_{\mu_1}
 (-k_{l2}\tau)
 + D_2 H^{(2)}_{\mu_2} (-k_{l2}\tau) \right),
  \end{align}
  where $H_{\mu}^{(i)}(x)$ ($i=1,2$) are  the first and second kinds of the Hankel functions and $C_{1,2},\, D_{1,2}$ are integration
constants.

\subsubsection{Matching condition}

As we discussed in the previous section, there are two phases in our model and we obtained perturbation modes for each phase separately.
Then all perturbed modes should satisfy
matching conditions at the transition point $\tau_{eq}$ in
conformal time,
\begin{align}\label{jcondition}
&{\cal V}_{e}(\tau)|_{\tau = \tau_{eq}} ={\cal V}_{l}(\tau)_{\tau = \tau_{eq}}, \quad {\cal V}_{e}'(\tau)_{\tau = \tau_{eq}} ={\cal V}_{l}'(\tau)_{\tau = \tau_{eq}},
\nn \\
&{\cal U}_{e}(\tau)_{\tau = \tau_{eq}} ={\cal U}_{l}(\tau)_{\tau = \tau_{eq}},
\quad {\cal U}_{e}'(\tau)_{\tau = \tau_{eq}} ={\cal U}_{l}'(\tau)_{\tau = \tau_{eq}}.
\end{align}
Here we notice that the perturbed modes ${\cal V}$ and ${\cal U}$ satisfy the same type of differential equations
with different parameters. So in what follows, we  only consider the matching condition
for the mode ${\cal V}$.
Then the  results can be extended to the case of the mode ${\cal U}$ as well.
From the matching condition in \eqref{jcondition} we obtain
the corresponding integration constants, \begin{align}
C_1-C_2 = &{e^{-i\beta\tau_{eq}}\sqrt{\pi}  \over  \sqrt{2}\, k\sqrt{-k_{e1}\tau_{eq}}}
\Biggr[(k_{e1}+i \alpha M_{{\rm P}})k_{l1}\tau_{eq}J_{\mu_1 -1} \nn \\
&~+\left(
\left(k_{e1}+i \alpha M_{{\rm P}}\right)\left(\mu-{1\over2}\right)
+\left(i \alpha^2 M_{{\rm P}}^2+2\alpha k_{e1}
M_{{\rm P}}-2ik_{e1}^2\right)\tau_{eq}
\right)J_{\mu_1}
\Biggr]
 \,, \nn \\
C_1+C_2 = &{-ie^{-i\beta\tau_{eq}}\sqrt{\pi}  \over  \sqrt{2} \,k\sqrt{-k_{e1}\tau_{eq}}}
\Biggr[(k_{e1}+i \alpha M_{{\rm P}})k_{l1}\tau_{eq}Y_{\mu_1 -1} \nn \\
&~+\left(
\left(k_{e1}+i \alpha M_{{\rm P}}\right)
\left(\mu-{1\over2}\right)
+\left(i \alpha^2 M_{{\rm P}}^2+2\alpha k_{e1}
M_{{\rm P}}-2ik_{e1}^2\right)\tau_{eq}
\right)Y_{\mu_1}
\Biggr]
 \,,
\end{align}
where we used the relations between the Hankel functions and  Bessel functions, $H^{(1,2)}_\mu(x) \equiv
J_\mu(x) \pm i Y_\mu (x),$ and
defined the quantities at the transition point as \begin{align}
&J_\mu = J_{\mu}\left(-k_{l1} \tau_{eq}\right) \,, \quad
Y_\mu = Y_{\mu}\left(-k_{l1} \tau_{eq}\right) \,, \quad
\tau_{eq} = -{\sqrt{2}\coth^{-1}(\sqrt{2}) \over
\alpha M_{{\rm P}}}.
\end{align}

\subsubsection{Power spectrum}

Now we try to obtain the power spectrum for the curvature perturbation ${\cal R}$ in \eqref{ICP}
and related observational quantities.
For a single scalar model, one usually reads the power spectrum at the horizon
crossing point since it is guaranteed in the absence of the transition point
that curvature perturbations of perturbed modes are frozen after the horizon
crossing. In our case  with two inflationary phases, however,
reading the power spectrum at the horizon crossing point can cause some possible errors
for large scale modes which are deformed due to the presence of the nonvanishing $\alpha$-term. That is, one can not guarantee  the freezing of the curvature perturbation after the horizon crossing for large scale modes. For this reason, we read the power spectrum at the limit  $\tau \to 0$ for all values
in the region $k>\alpha M_{{\rm P}}$.

Using the relation \eqref{ABapprox} in the limit $\frac{3\alpha^2M_{{\rm P}}^4}{2 a^2}\,
\ll\, V(\varphi) $, we obtain leading contributions for the curvature perturbation~\cite{Koh:2013msa},
\begin{align}\label{Rapprox}
{\cal R} \simeq
\frac1{2 + \frac{\alpha^2 M_{{\rm P}}^2}{a^2H^2\epsilon}}
\left( - \frac{\sqrt{2}}{M_{{\rm P}}
\sqrt{\epsilon}}\, Q_\varphi + \frac{\alpha}{k H\epsilon}\, \dot Q_u\right).
\end{align}
As we see in \eqref{kminval}, the comoving wave number
has a minimum value, and so
we notice that all comoving wave numbers relevant
to the observation are in the range
$\frac{\alpha M_{{\rm P}}}{k} <1$.
Due to this fact,
from now on we neglect the contribution of
$\dot Q_u$ in \eqref{Rapprox}
by keeping the leading contribution of $\frac{\alpha M_{{\rm P}}}{k}$ since
the $\dot Q_u$-term in \eqref{Rapprox} gives ${\cal O}\left(\frac{\alpha M_{{\rm P}}}{k}\right)^4$ contribution to the the resulting power
spectrum~\cite{Koh:2013msa}.
Then the power spectrum of the curvature perturbation is given by
\begin{align}\label{Papprox}
{\cal P}_{\cal R}(k) &\equiv \frac{k^3}{2\pi^2}
\langle {\cal R}{\cal R}^*\rangle_*
\nn \\
&\simeq
\left(1+{\alpha^2 M_{{\rm P}}^2 \over 2 \epsilon_* k^2} \right)^{-2}
{H_*^2\over 2 \epsilon_* M_{{\rm P}}^2}
\lim_{\tau\rightarrow 0}
k_{1l} \langle{\cal V}_{l}(\tau){\cal V}_{l}(\tau)^*\rangle,
\end{align}
where the subscripted asterisk indicates the value at the horizon crossing point
$k_{l1} = aH$ and we take the late time limit $\tau \rightarrow 0$ to read the
power spectrum.
Plugging the first line of \eqref{laterVUsol} into \eqref{Papprox}, we obtain
\begin{align}
{\cal P}_{\cal R} &\simeq
{\cal P}_{\cal R}^{(0)} {\big|C_1-C_2\big|^2 \over
\left(1+{\alpha^2 M_{{\rm P}}^2\over 2 \epsilon k^2}\right)^2}
\,,\label{repower}
\end{align}
where ${\cal P}_{\cal R}^{(0)}$ denotes the power spectrum of the canonical single inflation model,
\begin{align}
&{\cal P}_{\cal R}^{(0)} =  {H_*^2 \over 8 \pi^2 M_{{\rm P}}^2}{1\over \epsilon}
\Big(1+\left(2-2C\right)\eta+\left(6C-8\right)
\epsilon\Big)\,.
\end{align}
Here,
$C=2-\ln2 -\gamma$ with the
Euler-Mascheroni constant $\gamma \approx 0.5772$ and
\begin{align}
\big|C_1-C_2\big|^2 &=-
{ \pi \alpha M_{{\rm P}}\over Xk_{e1} }
\Biggr[ \beta^2\left(1 + \frac{\alpha^2 M_{{\rm P}}^2}{2 k^2}\right)
{k_{l1}^2 \over \alpha^2 M_{{\rm P}}^2}
J_{\mu_1-1}^2
\nn \\
&~~~~~~~~~~~~~~~~~+2\beta\left\{\left(1 + \frac{\alpha^2 M_{{\rm P}}^2}{2 k^2}\right)
\left(\mu-{1\over2}\right)+\frac{\beta\alpha^2 M_{{\rm P}}^2}{k^2}
\right\}{k_{l1} \over \alpha M_{{\rm P}}}J_{\mu_1}J_{\mu_1-1} \nn \\
&~~~~~~~~~~~~~~~~~~+\biggr\{ \left(1 + \frac{\alpha^2 M_{{\rm P}}^2}{2 k^2}\right)
\left(\mu-{1\over2}\right)^2
+\frac{\beta \alpha^2 M_{{\rm P}}^2}{ k^2}
\left(\mu-{1\over2}\right)
\nn \\
&~~~~~~~~~~~~~~~~~~+\beta^2\left(1+ \frac{3\alpha^2
M_{{\rm P}}^2}{4 k^2}\right)\biggr\}J_{\mu_1}^2
\Biggr],
\end{align}
where $\beta \equiv M_{{\rm P}} \alpha \tau_{eq} = -\sqrt{2}\coth^{-1}(\sqrt{2})$.
\begin{figure*}[ht]
\begin{center}
\scalebox{0.80}[0.80]{\includegraphics{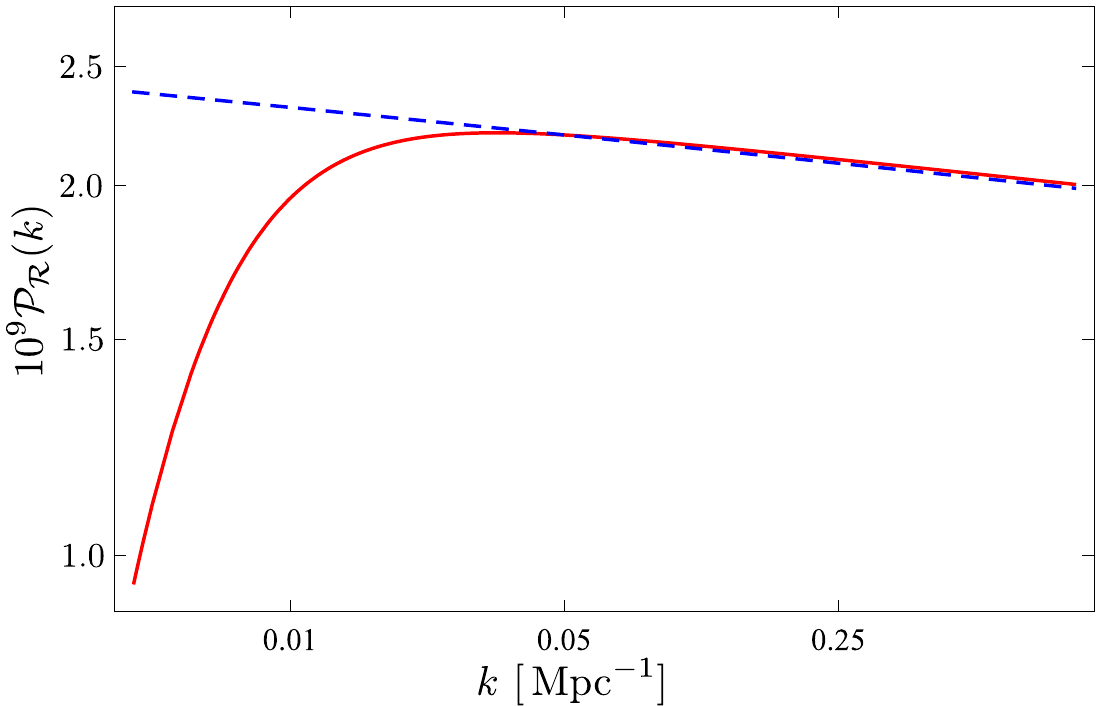}}\label{Fig2}
\end{center}
\caption{\small
The primordial power spectrum of
curvature perturbation for the usual single field model~(dashed blue line)
and the inflation model on the spatial condensation~(solid red line).
}\label{figpower}
\end{figure*}

As discussed previously, we notice that
during the early time phase,  modes satisfying the condition
$k > \alpha M_{{\rm P}}$ can only cross the Hubble horizon.
In other words,  modes in the range
$k < \alpha M_{{\rm P}}$ stay outside the Hubble horizon
and never cross the horizon. Therefore, those super horizon modes are causally
disconnected to our universe and irrelevant to observational quantities.
As we see the plot of power spectrum in Fig.2,
the power spectrum in our model asymptotically
approaches that of the single field model
(dashed blue line) by increasing the comoving wave
number $k$, while it is strongly suppressed
by decreasing the value $k$.

In this paper, we investigate the behaviour of
power spectrum in terms of the value
$k$. To do that, we divide the values of $k$ into two regions,
$k \gg \alpha M_{{\rm P}}$ and $k\gtrsim \alpha M_{{\rm P}}$.
At first in the region $k \gg \alpha M_{{\rm P}}$,
from \eqref{repower} we have the following asymptotic form
of power spectrum
\begin{align}
{\cal P_R}
&\simeq {\cal P}_{\cal R}^{(0)}
\Biggr[
1-{\alpha^2 M_{{\rm P}}^2\over \epsilon k^2}
+{\alpha^2 M_{{\rm P}}^2\over \beta^2 k^2}
\sin^2\left({\beta k \over \alpha M_{{\rm P}}}\right)
\Biggr].
\label{resultpps2}
\end{align}
The power spectrum is almost scale
invariant  but modulated with small oscillation.
This oscillation behaviour of power spectrum
was also reported in different inflation model
with phase transition~\cite{Joy:2007na,Hazra:2014jka}.
The corresponding spectral index $n_{{\cal R}}$
and the running of the spectral index $\alpha_{{\cal R}}$
at the pivot scale $k_0$ in the leading order of $\alpha /k$
with small slow-roll parameters are given by
\begin{align}
n_{\cal R}  \simeq n_{\cal R}^{(0)}
+ {2\alpha^2 M_{{\rm P}}^2 \over \epsilon_* k_0^2},
\qquad
\alpha_{\cal R} \simeq
-{4\alpha^2 M_{{\rm P}}^2 \over \epsilon_* k_0^2},
\end{align}
where $n_{\cal R}^{(0)}$ denotes the spectral
index at the pivot scale $k_0$ for the
single field inflation model.
We notice  that the spectral index is slightly
increasing and the running is negative and slightly
decreasing by decreasing the value $k$
due to the spatial condensation.
On the other hand for the region
$k\gtrsim \alpha M_{{\rm P}}$, we obtain
the behaviour of the power spectrum as
\begin{align}
{\cal P_R}
&\sim
{ k^4 \over \alpha^4 M_p^4}
{\cal P}_{\cal R}^{(0)} \,.
\label{resultpps1}
\end{align}
This behaviour was plotted in Fig.2 in terms of the red line in logarithmic scale of
the wave number.
Then the corresponding spectral index and its running
 are given by
\begin{align}\label{largest}
n_{\cal R}  \simeq 5 \,,
\qquad
\alpha_{\cal R} \simeq 0 \,.
\end{align}
From this behaviour of the power spectrum in the spatial condensation,
one can clearly notice a strong suppression
of the scalar power spectrum on large scales.
Since the spectral index is approaching a constant on these  large
scale limit, the running of the spectral index becomes almost zero.

We analysed the behaviour of the power spectrum
in terms of semi-analytic methods for large scales $k\gtrsim \alpha M_{{\rm P}}$
and small scales $k \gg \alpha M_{{\rm P}}$ in the previous paragraph.
However, as we see the numerical result shown in Fig.2,
there is a sudden transition of the power spectrum for intermediate scales between
$k\gtrsim \alpha M_{{\rm P}}$ and  $k \gg \alpha M_{{\rm P}}$.
Therefore, for this region, the spectral index is growing suddenly by decreasing the comoving
wave number,  i.e., the scalar power spectrum starts to be suppressed strongly,
and then one has large negative running of the spectral index 
in the intermediate region.

\section{Conclusion}

There are several models to accomplish the suppression. 
One common property of these models is that there exists
a phase transition of the background evolution and it is connected to
the slow-roll phase of the single field inflation model at late time.
In this paper, we showed that a deformed single field inflation model
in terms of {\it the spatial condensation} has a phase transition which is
similar to that of models in \cite{Hazra:2014aea,Bousso:2014jca}
and the suppression of scalar power spectrum on large scales.

We deformed a single field inflation model in terms of
three SO(3) symmetric moduli fields $\sigma^a$.
On the solution with constant gradient
$\sigma^a = \alpha x^a$,
the background evolution is equivalent to
that of the single inflation model with curvature term of the open universe.
During very early time, the background evolution is governed
by the curvature term but soon after the curvature term is rapidly decreased.
Then at the late time, the  evolution is governed by the
potential term of the single scalar field and asymptotically approaches that of
a single field inflation model.
This means that there exists a phase transition of the background evolution,
and so,  for an analytic approach
we divided the background evolution into two phases, the $\alpha$-term
dominant phase and the potential term dominant phase.

During the $\alpha$-term dominant phase, we assumed that the inflation started
with a very large value of the $\alpha$-term  (curvature term)
and then the e-folding could be accumulated very rapidly.
Therefore, during the early time phase with a short cosmic time process,
the single scalar field remains almost constant.
Assuming the scalar potential is a constant, there is an exact solution
governing the background evolution.
On the other hand, in the late time phase,
the $\alpha$-term becomes very small and the evolution is governed by
the potential term and asymptotically approaches that of the single field inflation.
Under the slow-roll assumption, the system is governed by
slow-roll parameters and small contribution of the $\alpha$-term.

Under the above circumstance of the background evolution, we investigated
the behaviour scalar modes in linear perturbation level.
We considered the perturbation modes in the early and late time phases separately.
For perturbed modes in the two phases, we applied the junction condition
at the transition point.
Then we obtained the power spectrum, the spectral index, and the running
of the spectral index for scalar modes.
We found that the power spectrum is apparently suppressed by decreasing
the comoving wave number, while it approaches the value of the single
field inflation model for large value of the comoving wave number.
Therefore, one can obtain large negative running of the scalar spectral index
on large scales. We also found a oscillation behaviour of the power spectrum at late time.

We focused on the suppression of the scalar power spectrum
on large scales. 
However, we also expect that there will be a nontrivial contribution
to the isocurvature perturbation since our model has two
perturbed scalar modes.
Actually  our model introduces a free gradient parameter $\alpha$
to a single field inflation model in isotropic and homogeneous way.
Therefore, in order to accommodate observational data,
similar analysis as we did in this paper can be applied
to various inflation models by adjusting the free parameter $\alpha$.

\section*{Acknowledgments}
This work was supported by the Korea Research Foundation Grant
funded by the World Class University grant no. R32-10130 (S.K.,O.K.),
the National Research
Foundation of Korea(NRF) grant funded by the Korea government(MEST)
through the Center for Quantum Space- time(CQUeST) of Sogang University
with grant number 2005-0049409 (P.O.),  the BSRP
through the NRF of Korea funded
by the MEST (No. 2010-210021996) (P.O.),  (NRF-2014R1A1A2059761)(O.K.), and  the Mid-career Researcher Program through the NRF grant funded by the Korean government (MEST) (No.
2014-051185) (O.K.).


\begin{thebibliography}{99}


\bibitem{Ade:2013zuv}
  P.~A.~R.~Ade {\it et al.}  [Planck Collaboration],
  ``Planck 2013 results. XVI. Cosmological parameters,''
  Astron.\ Astrophys.\  (2014)
  [arXiv:1303.5076 [astro-ph.CO]].


\bibitem{WMAP9}
  P.~A.~R.~Ade {\it et al.}  [WMAP Collaboration],
  ``Nine-Year Wilkinson Microwave Anisotropy Probe (WMAP) Observations: Cosmological Parameter Results,''
  Astrophys. J. Suppl. {\bf 208} (2013) 19   [arXiv:1212.5226 [astro-ph.CO]].
  
\bibitem{Hazra:2014jka}
 D.~K.~Hazra, A.~Shafieloo, G.~F.~Smoot and A.~A.~Starobinsky,
  ``Whipped inflation,''
  Phys.\ Rev.\ Lett.\  {\bf 113}, 071301 (2014)
  [arXiv:1404.0360 [astro-ph.CO]];

\bibitem{Joy:2007na}
  M.~Joy, V.~Sahni and A.~A.~Starobinsky,
  ``A New Universal Local Feature in the Inflationary Perturbation Spectrum,''
  Phys.\ Rev.\ D {\bf 77}, 023514 (2008)
  [arXiv:0711.1585 [astro-ph]].

\bibitem{Hazra:2014goa} 
  D.~K.~Hazra, A.~Shafieloo, G.~F.~Smoot and A.~A.~Starobinsky,
  ``Wiggly Whipped Inflation,''
  JCAP {\bf 1408}, 048 (2014)
  [arXiv:1405.2012 [astro-ph.CO]].

\bibitem{Lyth:1990dh} 
  D.~H.~Lyth and E.~D.~Stewart,
  ``Inflationary density perturbations with Omega $<$ 1,''
  Phys.\ Lett.\ B {\bf 252}, 336 (1990).

\bibitem{Ratra:1994vw} 
  B.~Ratra and P.~J.~E.~Peebles,
  ``Inflation in an open universe,''
  Phys.\ Rev.\ D {\bf 52}, 1837 (1995).

\bibitem{White:2014aua} 
  J.~White, Y.~l.~Zhang and M.~Sasaki,
  ``Scalar suppression on large scales in open inflation,''
  Phys.\ Rev.\ D {\bf 90}, no. 8, 083517 (2014)
  [arXiv:1407.5816 [astro-ph.CO]].









  
\bibitem{Ade:2014xna}
  P.~A.~R.~Ade {\it et al.}  [BICEP2 Collaboration],
  ``BICEP2 I: Detection Of B-mode Polarization at Degree Angular Scales,''
  arXiv:1403.3985 [astro-ph.CO].


\bibitem{Mortonson:2014bja}
  M.~J.~Mortonson and U.~Seljak,
  ``A joint analysis of Planck and BICEP2 B modes including dust polarization uncertainty,''
  arXiv:1405.5857 [astro-ph.CO];

  R.~Flauger, J.~C.~Hill and D.~N.~Spergel,
  ``Toward an Understanding of Foreground Emission in the BICEP2 Region,''
  arXiv:1405.7351 [astro-ph.CO].

\bibitem{Adam:2014bub}
  R.~Adam {\it et al.}  [Planck Collaboration],
  ``Planck intermediate results. XXX. The angular power spectrum of polarized dust emission at intermediate and high Galactic latitudes,''
  arXiv:1409.5738 [astro-ph.CO].

\bibitem{Cheng:2014pxa}
  C.~Cheng, Q.~G.~Huang and S.~Wang,
  ``Constraint on the primordial gravitational waves from the joint analysis of BICEP2 and Planck HFI 353 GHz dust polarization data,''
  arXiv:1409.7025 [astro-ph.CO].


\bibitem{Contaldi:2014zua}
  C.~R.~Contaldi, M.~Peloso and L.~Sorbo,
  ``Suppressing the impact of a high tensor-to-scalar ratio on the temperature anisotropies,''
  JCAP {\bf 1407}, 014 (2014)
  [arXiv:1403.4596 [astro-ph.CO]];
  


  V.~Miranda, W.~Hu and P.~Adshead,
  ``Steps to Reconcile Inflationary Tensor and Scalar Spectra,''
  arXiv:1403.5231 [astro-ph.CO];

  K.~N.~Abazajian, G.~Aslanyan, R.~Easther and L.~C.~Price,
  ``The Knotted Sky II: Does BICEP2 require a nontrivial primordial power spectrum?,''
  JCAP {\bf 1408}, 053 (2014)
  [arXiv:1403.5922 [astro-ph.CO]];
  
  A.~Ashoorioon, K.~Dimopoulos, M.~M.~Sheikh-Jabbari and G.~Shiu,
  ``Non-Bunch–Davis initial state reconciles chaotic models with BICEP and Planck,''
  Phys.\ Lett.\ B {\bf 737}, 98 (2014)
  [arXiv:1403.6099 [hep-th]];

  L.~Lello and D.~Boyanovsky,
  ``Tensor to scalar ratio and large scale power suppression from pre-slow roll initial conditions,''
  JCAP {\bf 1405}, 029 (2014)
  [arXiv:1312.4251 [astro-ph.CO]];
  


  H.~Firouzjahi and M.~H.~Namjoo,
  ``Jump in fluid properties of inflationary universe to reconcile scalar and tensor spectra,''
  arXiv:1404.2589 [astro-ph.CO];

  B.~Hu, J.~W.~Hu, Z.~K.~Guo and R.~G.~Cai,
  ``Reconstruction of the primordial power spectra with Planck and BICEP2,''
  Phys.\ Rev.\ D {\bf 90}, 023544 (2014)
  [arXiv:1404.3690 [astro-ph.CO]];

  C.~Cheng and Q.~G.~Huang,
  ``Probing the primordial Universe from the low-multipole CMB data,''
  arXiv:1405.0349 [astro-ph.CO];

  R.~Kallosh, A.~Linde and A.~Westphal,
  ``Chaotic Inflation in Supergravity after Planck and BICEP2,''
  Phys.\ Rev.\ D {\bf 90}, 023534 (2014)
  [arXiv:1405.0270 [hep-th]];

  Y.~Wan, S.~Li, M.~Li, T.~Qiu, Y.~Cai and X.~Zhang,
  ``Single field inflation with modulated potential in light of the Planck and BICEP2,''
  Phys.\ Rev.\ D {\bf 90}, 023537 (2014)
  [arXiv:1405.2784 [astro-ph.CO]];

  K.~Kohri and T.~Matsuda,
  ``Ambiguity in running spectral index with an extra light field during inflation,''
  arXiv:1405.6769 [astro-ph.CO];

  M.~W.~Hossain, R.~Myrzakulov, M.~Sami and E.~N.~Saridakis,
  ``Evading Lyth bound in models of quintessential inflation,''
  Phys.\ Lett.\ B {\bf 737}, 191 (2014)
  [arXiv:1405.7491 [gr-qc]];

  A.~Ashoorioon, C.~van de Bruck, P.~Millington and S.~Vu,
  ``Effect of transitions in the Planck mass during inflation on primordial power spectra,''
  Phys.\ Rev.\ D {\bf 90}, 103515 (2014)
  [arXiv:1406.5466 [astro-ph.CO]];

  M.~Cicoli, S.~Downes, B.~Dutta, F.~G.~Pedro and A.~Westphal,
  ``Just enough inflation: power spectrum modifications at large scales,''
  arXiv:1407.1048 [hep-th].

\bibitem{Hazra:2014aea}
  D.~K.~Hazra, A.~Shafieloo, G.~F.~Smoot and A.~A.~Starobinsky,
  ``Ruling out the power-law form of the scalar primordial spectrum,''
  JCAP {\bf 1406}, 061 (2014)
  [arXiv:1403.7786 [astro-ph.CO]].


\bibitem{Bousso:2014jca}
  R.~Bousso, D.~Harlow and L.~Senatore,
  ``Inflation After False Vacuum Decay: New Evidence from BICEP2,''
  arXiv:1404.2278 [astro-ph.CO];
  ``Inflation after False Vacuum Decay: Observational Prospects after Planck,''
  arXiv:1309.4060 [hep-th].



\bibitem{ArmendarizPicon:2007nr}
  C.~Armendariz-Picon,
  ``Creating Statistically Anisotropic and Inhomogeneous Perturbations,''
  JCAP {\bf 0709}, 014 (2007)
  [arXiv:0705.1167 [astro-ph]].

\bibitem{Lee:2009zv}
  J.~Lee, T.~H.~Lee, T.~Y.~Moon and P.~Oh,
  ``De-Sitter nonlinear sigma model and accelerating universe,''
   Phys.\ Rev.\ D {\bf 80}, 065016 (2009)  [arXiv:0905.2653 [gr-qc]].  

\bibitem{Endlich:2012pz}
  S.~Endlich, A.~Nicolis and J.~Wang,
  ``Solid Inflation,''
  JCAP {\bf 1310}, 011 (2013)
  [arXiv:1210.0569 [hep-th]].


\bibitem{Koh:2013msa}
  S.~Koh, S.~Kouwn, O-K.~Kwon and P.~Oh,
  ``Cosmological Perturbations of a Quartet of Scalar Fields with a Spatially Constant Gradient,''
  Phys.\ Rev.\ D {\bf 88}, 043523 (2013)
  [arXiv:1304.7924 [gr-qc]].



\bibitem{Bartolo:2013msa}
  N.~Bartolo, S.~Matarrese, M.~Peloso and A.~Ricciardone,
  ``Anisotropy in solid inflation,''
  JCAP {\bf 1308}, 022 (2013)
  [arXiv:1306.4160 [astro-ph.CO]];

  C.~Lin,
  ``Massive Graviton on a Spatial Condensate,''
  arXiv:1307.2574 [hep-th];

  A.~Nicolis, R.~Penco and R.~A.~Rosen,
  ``Relativistic Fluids, Superfluids, Solids and Supersolids from a Coset Construction,''
  Phys.\ Rev.\ D {\bf 89}, 045002 (2014)
  [arXiv:1307.0517 [hep-th]];

  S.~Endlich, B.~Horn, A.~Nicolis and J.~Wang,
  ``The squeezed limit of the solid inflation three-point function,''
  Phys.\ Rev.\ D {\bf 90}, 063506 (2014)
  [arXiv:1307.8114 [hep-th]];

  M.~Akhshik, R.~Emami, H.~Firouzjahi and Y.~Wang,
  ``Statistical Anisotropies in Gravitational Waves in Solid Inflation,''
  JCAP {\bf 1409}, 012 (2014)
  [arXiv:1405.4179 [astro-ph.CO]];

  N.~Bartolo, M.~Peloso, A.~Ricciardone and C.~Unal,
  ``The expected anisotropy in solid inflation,''
  arXiv:1407.8053 [astro-ph.CO];

  E.~Dimastrogiovanni, M.~Fasiello, D.~Jeong and M.~Kamionkowski,
  ``Inflationary tensor fossils in large-scale structure,''
  arXiv:1407.8204 [astro-ph.CO];

  M.~Akhshik,
  ``Clustering Fossils in Solid Inflation,''
  arXiv:1409.3004 [astro-ph.CO].

\bibitem{Linde:1998iw}
  A.~D.~Linde,
  ``A Toy model for open inflation,''
  Phys.\ Rev.\ D {\bf 59}, 023503 (1999)
  [hep-ph/9807493];

  B.~Freivogel, M.~Kleban, M.~Rodriguez Martinez and L.~Susskind,
  ``Observational consequences of a landscape,''
  JHEP {\bf 0603}, 039 (2006)
  [hep-th/0505232];

  A.~D.~Linde, M.~Sasaki and T.~Tanaka,
  ``CMB in open inflation,''
  Phys.\ Rev.\ D {\bf 59}, 123522 (1999)
  [astro-ph/9901135].



\bibitem{Masso:2006gv}
  E.~Masso, S.~Mohanty, A.~Nautiyal and G.~Zsembinszki,
  ``Imprint of spatial curvature on inflation power spectrum,''
  Phys.\ Rev.\ D {\bf 78}, 043534 (2008)
  [astro-ph/0609349].


\bibitem{Yamamoto:1995sw}
  K.~Yamamoto, M.~Sasaki and T.~Tanaka,
  ``Large angle CMB anisotropy in an open universe in the one bubble inflationary scenario,''
  Astrophys.\ J.\  {\bf 455}, 412 (1995)
  [astro-ph/9501109].


\end{thebibliography}
\end{document}